\documentclass[12pt,notitlepage]{revtex4-1}
\usepackage{amsmath}
\usepackage{amsfonts}
\usepackage{amssymb}
\usepackage{graphicx}
\usepackage{epstopdf}
\usepackage{bbm}

\usepackage{color}
\usepackage{xcolor}
\usepackage{dsfont}

scaled\magstep1

     \let\d=\delta 
       
\let\m=\mu                 \let\r=\rho
\let\s=\sigma     
   \let\o=\omega
   \let\L=\Lambda 
         
\let\O=\Omega

\def\\{\hfill\break}
\def\={:=}

\let\0=\noindent

\def\tende#1{\,\vtop{\ialign{##\crcr\rightarrowfill\crcr\noalign{\kern-1pt
    \nointerlineskip} \hskip3.pt${\scriptstyle #1}$\hskip3.pt\crcr}}\,}
\def\otto{\,{\kern-1.truept\leftarrow\kern-5.truept\to\kern-1.truept}\,}

\def\to{\rightarrow}

\def\qed{\hfill\raise1pt\hbox{\vrule height5pt width5pt depth0pt}}

\def\be{\begin{equation}}
\def\ee{\end{equation}}
\def\bea{\begin{eqnarray}}
\def\eea{\end{eqnarray}}
\def\nn{\nonumber}

\def\Tr{\mathrm{Tr}}

\newtheorem{theorem}{Theorem}[section]

\numberwithin{equation}{section}

\begin{document}

\title{Universality of charge transport in weakly interacting fermionic systems}

\author{Alessandro Giuliani}
\affiliation{University of Roma Tre, Department of Mathematics and Physics, L.go S. L. Murialdo 1, 00146 Roma, Italy}
\author{Vieri Mastropietro}
\affiliation{University of Milano, Department of Mathematics ``F. Enriquez'',\\
 Via C. Saldini 50, 20133 Milano, Italy}
\author{Marcello Porta}
\email{marcello.porta@math.uzh.ch}
\affiliation{University of Z\"urich, Mathematics Department, Winterthurerstrasse 190, 8057 Z\"urich, Switzerland}




\begin{abstract}
We review two rigorous results on the transport properties of weakly interacting fermionic systems on $2d$ lattices, in the linear response regime. First, we discuss the universality of the longitudinal conductivity for interacting graphene. Then, we focus on the transverse conductivity of general weakly interacting gapped fermionic systems, and we establish its universality. This last result proves the stability of the integer quantum Hall effect against weak interactions. The proofs are based on combinations of fermionic cluster expansion techniques, renormalization group and lattice Ward identities.
\end{abstract}

\maketitle


\section{Introduction}

Two-dimensional condensed matter systems often present remarkable transport properties. A paradigmatic phenomenon is the {\it integer quantum Hall effect} (IQHE): when exposed to strong transverse magnetic fields, ultrathin samples of suitable materials display a {\it quantized} transverse (or Hall) conductivity, $\s_{12}$. Namely, $\s_{12}$ only takes integer values, in units of $\frac{e^{2}}{h}$ with $e$ the electric charge and $h$ Planck's constant. This phenomenon was experimentially discovered in  \cite{Klitzing}, and elicited enormous interest in the theoretical physics community.

By now, the understanding of the IQHE for {\it noninteracting particles} (with or without disorder) reached a deep mathematical level \cite{AS2, Bellissard, AS2index, AG}. For translation invariant lattice models the quantization of $\s_{12}$ has a beautiful topological explanation: the Hall conductivity, as given by Kubo formula, is equal to the first Chern number of a suitable bundle (the {\it Bloch bundle}), defined starting from the Bloch functions of the system \cite{AS2}. Later, this result has been extended to disordered systems, where the lack of translation invariance makes the analysis much harder. This was first done in \cite{Bellissard} using noncommutative geometric methods, and then in  \cite{AS2index} using a functional analytic approach (see also  \cite{AG}). 

\medskip

Another interesting example of charge transport in $2d$ is provided by {\it graphene}. Graphene is a recently discovered material, that consists of a single monoatomic layer of graphite. It is the first realization of a stable (quasi-)$2d$ crystal, and has unique structural and electronic properties. For instance, at low temperature its longitudinal conductivity $\sigma_{11}$ appears to be equal to $\frac{e^{2}}{h}\frac{\pi}{2}$, with very high precision \cite{Na}. That is, $\sigma_{11}$ {\it only depends on fundamental constants}; it does not appear to be sensitive to lattice details, or to interactions among charge carriers.

From a theoretical viewpoint, this experimental observation agrees with the explicit computation of $\s_{11}$ performed starting from the Kubo formula for a {\it noninteracting} lattice model \cite{SPG}. Remarkably, the result is the {\it same} as obtained for the continuum (or ``relativistic'') approximation of the lattice model, which is obtained by linearizing the energy-momentum dispersion relation close to the Fermi energy.

\medskip

Many-body interactions are unavoidably present in real samples, and it is therefore important to understand their effect from a theoretical viewpoint. Concerning the IQHE \cite{OpenProblems}, a mechanism for quantization in the presence of interactions has been proposed in \cite{FrK, FrST}. It is based on the study of the effective actions of nonrelativistic many-body quantum systems, in the limit of low frequencies and long wavelengths. An important assumption in this approach is the {\it incompressibility} of the system: the interacting ground state is separated by the rest of the spectrum by a gap. From a mathematical viewpoint, the quantization of $\s_{12}$ has been recently proven in  \cite{HM}, for interacting particles on a torus. The proof holds for incompressible systems with nondegenerate ground state. These assumptions, however, are unproven in most of the physically relevant cases; exceptions are quantum perturbations of classical systems \cite{DFF, DFFRB}, and ``frustration free'' systems \cite{BHM}.

Concerning graphene, the effect of many-body interactions on the optical conductivity generated a debate in the theoretical physics literature \cite{He, Mi}. Analyses based on lowest order perturbation theory and on effective continuum models gave rise to predictions that appeared to be sensitive to the details of the regularization schemes, and therefore were far from providing a conclusive explanation of the universality of $\s_{11}$.

\medskip

Here we review two theorems on the transport properties of interacting $2d$ lattice models. In Section \ref{sec:graph} we prove the universality of the longitudinal conductivity of graphene, in the presence of weak many-body interactions. Then, in Section \ref{sec:IQHE} we prove the universality of the transverse conductivity for general weakly interacting gapped fermionic systems. This proves, in particular, the stability of the IQHE against weak interactions. The methods used are combinations of rigorous renormalization group (RG), fermionic cluster expansion and lattice Ward identities.

\section{Universality of the optical conductivity in graphene}\label{sec:graph}

Let $\L_{L}$ be the triangular lattice:
\be\label{eq:deflat}
\L_{L} = \{ \vec x \mid \vec x = n_{1}\ell_{1} + n_{2}\ell_{2},\;\quad n_{i}\in \mathbb{Z},\;\quad 0\leq n_{i} \leq L-1 \}\;
\ee
with $\vec \ell_{1} = \frac{1}{2}(3, -\sqrt{3})$, $\vec \ell_{2} = \frac{1}{2}(3, \sqrt{3})$. We assume periodic boundary conditions. The honeycomb lattice of graphene can be thought as the superposition of two triangular sublattices, $\Lambda^{(A)}_{L} \equiv \L_{L}$ and $\Lambda^{(B)}_{L}\equiv \L^{(A)}_{L} + \vec\d_{1}$ with $\d_{1} = (1,\, 0)$. With each site $\vec x\in \L_{L}$ we associate fermionic creation/annihilation operators, $\psi^{\pm}_{\vec x,\rho,\s}$, with $\r\in \{A,B\}$ the sublattice label and $\s\in \{\uparrow,\downarrow\}$ the spin label. The fermionic operators satisfy the usual canonical anticommutation relations. The Fock space Hamiltonian reads:
\bea
\mathcal{H}_{L} &=& \mathcal{H}_{L}^{(0)} + U\mathcal{V}_{L}\;,\nn\\
\mathcal{H}_{L}^{(0)} &=& -t\sum_{ \substack{ \vec x\in \L_{L} \\ \s = \uparrow\downarrow} } \big[ \psi^{+}_{\vec x, A,\s}\psi^{-}_{\vec x, B,\s} + \psi^{+}_{\vec x, A,\s}\psi^{-}_{\vec x - \vec\ell_{1}, B,\s} + \psi^{+}_{\vec x, A,\s}\psi^{-}_{\vec x - \vec\ell_{2}, B,\s} + h.c.\big]\;,\nn\\
\mathcal{V}_{L} &=& \sum_{\vec x \in \L_{L}}\sum_{\rho\in \{A,B\}} \Big[ n_{\vec x,\rho, \uparrow} - \frac{1}{2}\Big] \Big[ n_{\vec x,\rho,\downarrow}  - \frac{1}{2}\Big]\;,
\eea
with $n_{\vec x,\rho,\s} = \psi^{+}_{\vec x, \rho,\s}\psi^{-}_{\vec x, \r,\s}$ the density operator, and $t>0$ the hopping parameter. $\mathcal{H}_{L}$ is the Hamiltonian of the Hubbard model on the honeycomb lattice \cite{GM}: $\mathcal{H}_{L}^{(0)}$ describes the nearest-neighbour hopping, and $\mathcal{V}_{L}$ is an on-site density-density interaction. The factors $-1/2$ ensure that $\mathcal{H}_{L}$ is hole-particle symmetric. The single-particle Hamiltonian reads, in the infinite volume limit:
\be
H^{(0)}_{\infty} = \int_{\mathcal{B}} d\vec k\, \hat H^{(0)}(\vec k)\;,\qquad \hat H^{(0)}(\vec k) = \begin{pmatrix} 0 & -t\Omega(\vec k)^{*} \\ -t\Omega(\vec k) & 0 \end{pmatrix}\;,
\ee
where $\Omega(\vec k) = 1 + e^{-i\vec k\cdot \vec\ell_{1}} + e^{-i\vec k\cdot \vec\ell_{2}}$ vanishes on only two points on the Brillouin zone $\mathcal{B}\simeq \mathbb{T}^{2}$, the Fermi points $k_{F}^{\pm} = \big( \frac{2\pi}{3},\, \pm \frac{2\pi}{3\sqrt{3}} \big)$. There, $\O(\vec k' + k_{F}^{\pm}) = ik'_{1} \pm k'_{2} + O(\vec k'^{2})$, which gives rise to an asymptotic ``relativistic'' dispersion close to $\vec k_{F}^{\pm}$. The Gibbs state associated with this model is:
\be
\langle \cdot \rangle_{\beta,L} = \frac{\Tr_{\mathcal{F}_{L}} \cdot e^{-\beta (\mathcal{H}_{L} - \mu\mathcal{N}_{L})}}{Z_{\beta,L}}\;,\qquad Z_{\beta,L} = \Tr_{\mathcal{F}_{L}}\, e^{-\beta (\mathcal{H}_{L} - \mu\mathcal{N}_{L})}\;,
\ee
where: the trace is over the fermionic Fock space $\mathcal{F}_{L}$, $\mu$ is the chemical potential and $\mathcal{N}_{L}$ is the number operator. We shall fix $\mu = 0$: this imposes the {\it half-filling condition} (thanks to the hole-particle symmetry of $\mathcal{H}_{L}$). In this situation, the Fermi surface is given by only two points, $k_{F}^{\pm}$.

We define the current operator as:
\bea\label{eq:current}
&&\vec J := \\
&& it\sum_{\substack{\vec x\in \L_{L} \\ \s=\uparrow\downarrow} }\Big[ \vec\d_{1}\psi^{+}_{\vec x, A, \s}\psi^{-}_{\vec x, B, \s} + \vec\d_{2}\psi^{+}_{\vec x,A, \s}\psi^{-}_{\vec x - \vec\ell_{1},B, \s} + \vec\d_{3}\psi^{+}_{\vec x,A, \s}\psi^{-}_{\vec x - \vec\ell_{2},B, \s} - h.c.\Big]\nn
\eea
where $\vec \d_{1} = (1,\, 0)$, $\vec\d_{2} = (1/2)\big( -1,\, \sqrt{3} \big)$, $\vec \d_{3} = (1/2)\big( -1,\, -\sqrt{3} \big)$ are the vectors connecting each site on the $A$-sublattice with the three nearest neighbours on the $B$-sublattice. Notice that in the infinite volume limit the current (\ref{eq:current}) is simply $i\big[ \mathcal{H}_{\infty}, \vec X \big]$, with $\vec X$ the second quantization of the position operator. 

For $t\in [0,\beta)$, let $\vec J(t) := e^{t \mathcal{H}_{L}} J e^{-t\mathcal{H}_{L}}$ be the imaginary-time evolution of $\vec J$, and $\hat J(\omega) := \int_{0}^{\beta} dt\, e^{i\omega t} \vec J(t)$ with Matsubara frequency $\omega\in (2\pi / \beta)\mathbb{Z}$. The conductivity matrix is defined according to Kubo formula \cite{Ku57}:
\be\label{eq:defcond}
\sigma_{ij} := - \lim_{\omega\to 0^{+}} \frac{1}{\omega}\Big[ \hat K_{ij}(\omega) - \hat K_{ij}(0) \Big]\;,
\ee
where $\hat K_{ij}(\omega)$ is the ground-state current-current correlation function:
\be
\hat K_{ij}(\omega) = \lim_{\beta,L\to\infty} \frac{1}{A \beta L^{2}}\big\langle {\bf T}\, \hat J_{i}(\omega) \hat J_{j}(-\omega) \big\rangle_{\beta,L}\;,
\ee
with ${\bf T}$ the fermionic time-ordering and $A = 3\sqrt{3}/2$ the area of the fundamental cell on the honeycomb lattice. Eq. (\ref{eq:defcond}) can be understood as the analytic continuation to imaginary times of the first term in the adiabatic expansion of the expectation of $\vec J$ in the presence of a weak, slowly varying external field (see next section). The following result \cite{GMP1, GMP2} gives a rigorous justification for the experimentally observed universality of the optical conductivity of graphene  \cite{Na}. 
\begin{theorem}{\bf [Universality of conductivity for interacting graphene.]}
The exists $U_{0}>0$ such that for $|U| < U_{0}$ the conductivity matrix is analytic in the interaction $U$, and it is given by (restoring $e$ and $h$):
\be\label{eq:univ}
\s_{11} = \s_{22} = \frac{e^{2}}{h}\frac{\pi}{2}\;,\qquad \s_{12} = -\s_{21} = 0\;.
\ee
\end{theorem}

{\bf Remarks.}
\begin{itemize}

\item[(i)] Notice that if $\hat K_{ij}(\omega)$ was differentiable in $\omega$ then $\s_{ii}$ would be zero, simply because $\hat K_{ii}(\omega) = \hat K_{ii}(-\omega)$ and the limit in Eq. (\ref{eq:defcond}) would reconstruct the derivative at $\omega = 0$. Instead, $\hat K_{ij}(\omega)$ is only continuous in $\omega$; as one can easily check already in the absence of interactions, replacing the limit in Eq. (\ref{eq:univ}) with the derivative would give rise to a logarithmic divergence. 

\item[(ii)] The proof relies on a combination of multiscale analysis, fermionic cluster expansion and RG. Rigorous RG methods have been successfully applied to condensed matter physics in the last 30 years, starting from  \cite{BG, FT} (see  \cite{Mbook} for a review). Recently, these methods have been used to prove the analyticity of the ground-state correlation functions for the Hubbard model on the honeycomb lattice \cite{GM}. The main feature of this model with respect to usual $2d$ systems is the point-like nature of the Fermi surface, together with the vanishing of the density of states at the Fermi points; these facts ultimately imply that the interaction is {\it irrelevant} in the RG sense. 

\item[(iii)] The key ingredient in the proof of the universality of conductivity is gauge invariance. {\it Lattice Ward identities} allow to detect a crucial cancellation between the renormalized parameters of the emergent relativistic theory, which arises in the RG flow. A similar strategy has been used to prove universal scaling relations in $1d$ systems \cite{BFM0, BFM1, Mbook}.
\end{itemize}

\section{Integer quantum Hall effect for weakly interacting systems}\label{sec:IQHE}

Let $\Lambda_{L}$ be a $2d$ periodic Bravais lattice, as in Eq. (\ref{eq:deflat}) but with general linearly independent basis vectors $\vec\ell_{1}$, $\vec \ell_{2}$. We consider interacting Hamiltonians of the form:
\bea\label{eq:gappedmodel}
&&\mathcal{H}_{L} = \mathcal{H}_{L}^{(0)} + U\mathcal{V}_{L}\;,\nn\\
&&\mathcal{H}_{L}^{(0)} = \sum_{\substack{\vec x,\vec y \in \L_{L} \\ \s,\s'\in I}} \psi^{+}_{\vec x,\s}H_{\s\s'}^{(0)}(\vec x - \vec y)\psi^{-}_{\vec y,\s'}\;,\quad \mathcal{V}_{L}^{(0)} = \sum_{\substack{ \vec x,\vec y\in \L_{L} \\ \s,\s'\in I }} n_{\vec x,\s}v_{\s\s'}(\vec x - \vec y) n_{\vec y,\s'}\;,
\eea
where $\psi^{\pm}_{\vec x,\s}$ are fermionic operators, with $\s\in I \simeq [1,\ldots, N] $ the label of an internal degree of freedom ({\it e.g.} spin, sublattice label), and $H^{(0)}_{\s\s'}(\vec x)$, $v_{\s\s'}(\vec x)$ are periodic functions on $\L_{L}$. We assume that $\overline{H^{(0)}_{\s\s'}(\vec x)} = H^{(0)}_{\s'\s}(-\vec x)$ (so that $\mathcal{H}^{(0)}_{L}$ is self-adjoint), and that $H^{(0)}(\vec x),\, v(\vec x)$ decay faster than any power in $\| \vec x \|_{L}$, with $\|\cdot \|_{L}$ the distance on the torus. Also, we assume that the Bloch Hamiltonian $\hat H^{(0)}(\vec k) = \sum_{\vec x\in \L_{L}} e^{i\vec k\cdot \vec x} H^{(0)}(\vec x)$ is {\it gapped}, and we put the Fermi energy $\m$ in the gap:
\be\label{eq:gapcond}
\delta_{\mu} := \inf_{\vec k \in \mathcal{B}}\text{dist} (\mu, \sigma(H^{(0)}(\vec k))) > 0\;.
\ee
For $U=0$ the conductivity matrix, as given by Kubo formula, has the form \cite{TKNN, AS2}:
\be
\s_{11} = \s_{22} = 0 \;,\qquad \sigma_{12} = -\sigma_{21} \in \frac{e^{2}}{h}\mathbb{Z}\;. 
\ee
The quantization of $\s_{12}$ has a topological interpretation \cite{AS2}; in particular, the value of $\s_{12}$ is stable against perturbations that do not close the spectral gap. It is natural to ask whether stability holds also against weak many-body interactions. The following theorem proves that this is indeed the case \cite{GMP3}.

\begin{theorem}{\bf [Stability of IQHE for weakly interacting systems.]}\label{thm:IQHE} The exists $U_{0}>0$ such that for $|U| < U_{0}$ the conductivity matrix is analytic in the interaction $U$, and it is given by:
\be
\s_{ij} = \sigma_{ij}|_{U=0}\;,\qquad i,j = 1,2\;.
\ee 
In particular (restoring $e$ and $h$) $\s_{ii} = 0$ and $\s_{12} = -\s_{21} \in (e^{2}/h)\mathbb{Z}$. 
\end{theorem}
{\bf Remarks.}
\begin{itemize} 
\item[(i)] The preliminary step in the proof is to construct the Gibbs state, in the $\beta, L\to\infty$ limit, for $U$ small. This is done via standard fermionic cluster expansion and determinant bounds \cite{BF, Brydges, GeM}. As a result, the Gibbs state is analytic for $|U|< U_{0}$, with $U_{0}\to 0$ as $\delta_{\m}\to 0$. In specific cases the estimate on $U_{0}$ could be improved using RG; we will not discuss this issue here.
\item[(ii)] Universality follows from a rigorous, nonperturbative formulation of some key ideas of  \cite{CH, IM}. It turns out that Ward identities can be used to show that all the contributions of order $n\geq 1$ in $U$ to $\hat K_{ij}(\omega)$ are {\it quadratic} in $\omega$ (plus a constant part). This, together with the analyticity in $U$ of $\hat K_{ij}(\omega)$ and its differentiability in $\omega$ (also implied by (\ref{eq:gapcond})) gives the desired result.
\end{itemize}

Finally, in the present case it is also possible to reconstruct the real-time Kubo formula starting from (\ref{eq:defcond}); that is, the {\it Wick rotation} can be rigorously proved \cite{GMP3}. We set $\L\equiv \L_{\infty}$, $\mathcal{H}\equiv \mathcal{H}_{\infty}$, and $X_{i} := \sum_{\vec x\in \L}\sum_{\s\in I} x_{i} n_{\vec x,\s}$.

\begin{theorem}{\bf [Wick rotation.]} Let $U\in (-U_{0},\, U_{0})$, with $U_{0}$ as in Theorem \ref{thm:IQHE}. Then, the following identity holds:
\be\label{eq:wick}
\s_{ij} = \lim_{\omega\to 0^{+}} \frac{1}{\o }\Big( i\int_{-\infty}^{0} dt\, e^{\o t}\, \langle\!\!
\langle \big[e^{i\mathcal{H}t} J_{i}e^{-i\mathcal{H}t}, J_{j}\big] \rangle\!\!\rangle_{\infty} -\langle
\!\!\langle\big[[\mathcal H,X_i],X_j\big]\rangle\!\!\rangle_\infty\Big)\;,
\ee
where $\langle\!\! \langle \cdot \rangle\!\!\rangle_\infty = \lim_{\beta\to\infty}\lim_{L\to\infty} (AL^{2})^{-1} \langle \cdot \rangle_{\beta,L}$ with $A = |\vec\ell_{1}\wedge \vec\ell_{2}|$.
\end{theorem}

{\bf Remark.}
\begin{itemize}
\item[(i)] Eq. (\ref{eq:wick}) is the first order in the adiabatic expansion for the average current, after introducing a weak, slowly varying time-dependent external field \cite{AG}.
\item[(ii)] The proof is based on Cauchy's and Vitali's theorems for analytic functions, on suitable bounds for correlations at {\it complex} times and on {\it Lieb-Robinson bounds} (LR); in particular, the construction of the Gibbs state together with LR bounds allows to prove the existence of the correlations at real times, as $\beta,L\to\infty$.
\end{itemize}

{\bf Acknowledgements.} 
We gratefully acknowledge financial support from: the A*MIDEX project (n$^o$
ANR-11-IDEX-0001-02) funded by the ``Investissements d'Avenir'' French
Government program, managed by the French National Research Agency (A.G.);
the PRIN National Grant {\it Geometric and analytic theory of Hamiltonian systems in finite and infinite dimensions} (A.G. and V. M.); the NCCR SwissMAP (M.P.).

\end{document}